\documentclass[twocolumn,prl,superscriptaddress]{revtex4}

\usepackage{amsmath, amsthm, amssymb,amsthm}
\usepackage{graphicx}

\def\be{\begin{equation}}
\def\ee{\end{equation}}

\def\ev#1{\langle#1\rangle}
\def\jd{q}

\newtheorem*{lemma}{Lemma}

\newcommand{\bm}{\mathbf}

\begin{document}

\title{Guess your neighbour's input:\\ a multipartite non-local game with no
quantum advantage}
\author{Mafalda L. Almeida}
\affiliation{ICFO-Institut de Ciencies Fotoniques, E--08860
Castelldefels, Barcelona, Spain}
\author{Jean-Daniel Bancal}
\affiliation{GAP-Optique, Universit\'e de Gen\`eve, CH--1211
Geneva, Switzerland}
\author{Nicolas Brunner}
\affiliation{H.H. Wills Physics Laboratory, University of Bristol, Bristol, BS8 1TL, United Kingdom}
\author{Antonio Ac\'\i n}
\affiliation{ICFO-Institut de Ciencies Fotoniques, E--08860
Castelldefels, Barcelona, Spain} \affiliation{ICREA-Instituci\'o
Catalana de Recerca i Estudis Avan\c{c}ats,
E--08010 Barcelona, Spain}
\author{Nicolas Gisin}
\affiliation{GAP-Optique, Universit\'e de Gen\`eve, CH--1211
Geneva, Switzerland}
\author{Stefano Pironio}
\affiliation{Laboratoire d'Information Quantique, Universit\'e Libre de Bruxelles, Belgium}

\begin{abstract}
We present a multipartite nonlocal game in which each player must
guess the input received by his neighbour. We show that quantum
correlations do not perform better than classical ones at this
game, for any prior distribution of the inputs. There exist,
however, input distributions for which general no-signalling
correlations can outperform classical and quantum correlations.
Some of the Bell inequalities associated to our construction
correspond to facets of the local polytope. Thus our multipartite
game identifies parts of the boundary between quantum and
post-quantum correlations of maximal dimension. These results
suggest that quantum correlations might obey a generalization of
the usual no-signalling conditions in a multipartite setting.
\end{abstract}

\maketitle In  recent years, the study and understanding of
quantum nonlocality -- the fact that certain quantum correlations
violate Bell inequalities \cite{Bell} -- has benefited from a
cross-fertilization with information concepts.

On one hand, nonlocality has been identified as a key resource for
quantum information processing. It allows, for instance, the
reduction of communication complexity \cite{qcreview}, and in the
device-independent scenario, where one wants to achieve an
information task without any assumption on the devices used in the
protocol, it can be exploited for secure key
distribution~\cite{diqkd}, state tomography \cite{st}, and
randomness generation~\cite{randomness}.

On the other hand,  information concepts have provided a deeper
understanding of the nature of quantum nonlocality. It is known in
particular that the no-signalling principle (no arbitrarily fast
communication between remote parties) is compatible with the
existence of correlations more nonlocal than those allowed in
quantum theory~\cite{PR,barrett}. However, recent works have shown that the
existence of such stronger-than-quantum correlations would have
deep information-theoretic consequences: they would, for instance,
collapse communication complexity \cite{CC} and allow  perfect
nonlocal computation \cite{NLC}. In a related direction, it has
been realized that quantum correlations actually obey a
strengthened version of no-signalling, the principle of
information causality~\cite{IC}.

Up to now, such questions  have been almost exclusively considered
in the bipartite scenario. Here our aim is to investigate the
separation between quantum and no-signalling correlations in a
multipartite scenario. For this, we introduce and study a simple
multipartite nonlocal game, \emph{Guess Your Neighbour's Input}
(GYNI).

In GYNI, $N$ distant players are arranged on a ring and each
receive an input bit $x_i\in\{0,1\}$ (see Fig.~1).
The goal is that each participant provides an output bit $a_i\in\{0,1\}$ equal to its right neighbour's input bit: 
\begin{equation}\label{out=in+1}
a_i=x_{i+1}\qquad \text{for all } i=1,\ldots, N ,
\end{equation}
where $x_{N+1}\equiv x_1$. The $2^N$ possible input strings
$\bm{x}=(x_1,\ldots,x_N)$  are chosen according to some  prior
distribution $\jd(\bm{x})=\jd(x_1,\ldots,x_N)$, which is known to
the parties. The figure of merit of the game is given by the
average winning probability
\begin{equation}\label{nplayer cit}
\omega= \sum_{\bm{x}} \jd(\bm{x}) P(\bm{a}_i=\bm{x}_{i+1}|\bm{x})\,,
\end{equation}
where
$P(\bm{a}_i=\bm{x}_{i+1}|\bm{x})=P(a_1=x_2,\ldots,a_N=x_1|x_1,\ldots,x_N)$
denotes the probability of obtaining the correct outputs
(\ref{out=in+1}) when the players have received the input string
$\bm{x}$. Of course, players are not allowed to communicate after
the inputs are distributed. Thus, their performance depends only
on the initially agreed strategy and on the shared physical
resources.

\begin{figure}
  \includegraphics[height=0.3\columnwidth,trim=200 250 200 200]{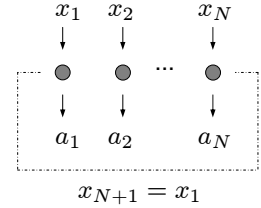}
  \caption{Representation of the GYNI nonlocal game. The goal is that each party outputs its right-neighbour's input: $a_i=x_{i+1}$.}
\end{figure}

The GYNI game captures a particular  notion of signalling: if the
players were able to win with high probability, their output would
reveal some information about their neighbour's input. We
therefore expect that the nonlocal correlations of quantum theory
cannot be exploited by non-communicating observers to perform
better at GYNI than using classical resources alone. We confirm
this intuition and prove that, indeed, quantum correlations
provide no advantage over classical correlations. Surprisingly,
however, the no-signalling principle is not at the origin of the
quantum limitation: for $N\geq3$, there exist input distributions
$\jd$ for which no-signalling correlations provide an advantage
over the best classical and quantum strategies. This suggests the
possibility that in a multipartite scenario, quantum correlations
obey a qualitatively stronger version of the usual no-signalling
conditions.

Each of the input distributions $\jd$ associated  with a
non-trivial no-signalling strategy defines a Bell inequality whose
maximal classical and quantum values coincide, but whose
no-signaling value is strictly larger. Interestingly, some of
these inequalities define facets of the polytope of local
correlations. We thus prove the existence of non-trivial facet
Bell inequalities with no quantum violation, answering a question
raised by Gill \cite{Prob26Werner}. Moreover, since these Bell
inequalities are facets, the GYNI game identifies a portion of the
boundary of the set of quantum correlations of non-zero measure,
in contrast with previous information-theoretic or physical
limitations on nonlocality \cite{CC,NLC,IC,emergence,allcock,ML}.

\textbf{GYNI with classical and quantum resources.}
We start by showing that the optimal classical and quantum winning strategies are identical for any prior distribution $\jd$ of the inputs.
Let us first show that there is a simple classical strategy achieving a winning probability
\be\label{cstrat}
\omega_c=\max_\bm{x}\left[\jd(\bm{x})+\jd(\bar{\bm{x}})\right]\,,
\ee
where $\bar{\bm{x}}$ denotes the ``negation" of the input string $\bm{x}$, $\bar{\bm{x}}=(\bar x_1,\ldots,\bar x_N)$ with $\bar x_i=x_i \oplus 1$, and $\oplus$ denotes addition modulo 2.
This strategy is based on the following simple observation.\\[0.5em]
\addtocounter{equation}{1}
\hspace*{2em}\parbox[top][][c]{0.8\columnwidth}{
Let $\bm{y}$ be an arbitrary string. If $\bm{x}\neq\bm{y},\bar{\bm{y}}$,\\ there exists an $i$ s.t. $x_i=y_i$ and $x_{i+1}\neq y_{i+1}$.}\hfill $(\theequation)$\\[0,5em]
Indeed, if this was not the case, we would have that for any  $i$,
either $x_i\neq y_i$ or $x_{i+1}=y_{i+1}$. But this would in turn
imply that either $\bm{x}=\bm{y}$ or $\bm{x}=\bar{\bm{y}}$, in
contradiction with the hypothesis.

Consider now a classical strategy specified by the string
$\bm{y}$, where each party outputs the bit $a_i=y_{i+1}$ if it
received the input $y_i$, and outputs $a_i=\bar y_{i+1}$ if it
received $\bar y_i$. It obviously follows that
$P(\bm{a}_i=\bm{y}_{i+1}|\bm{y})=1$ and
$P(\bm{a}_i=\bar{\bm{y}}_{i+1}|\bar{\bm{y}})=1$. On the other
hand, $P(\bm{a}_i=\bm{x}_{i+1}|\bm{x})=0$ for all
$\bm{x}\neq\bm{y},\bar{\bm{y}}$. Indeed, from observation
(\theequation), there exists an $i$ such that $x_{i}=y_{i}$, but
for which $a_i=y_{i+1}\neq x_{i+1}$. The winning probability of
this classical strategy is thus equal to
$\omega=q(\bm{y})+q(\bar{\bm{y}})$, which yields (\ref{cstrat}) if
we take $\bm{y}$ to be
$q(\bm{y})+q(\bar{\bm{y}})=\max_\bm{x}[q(\bm{x})+q(\bar{\bm{x}})]$.

We now prove that there is no better quantum (hence classical)
strategy. In the most general quantum protocol, the parties share
an entangled state $|\psi\rangle$ and perform projective
measurements on their subsystem dependent on their inputs $x_i$.
They then output their measurement results $a_i$. Denoting
$M_{a_i}^{x_i}$ the projection operator associated to the output
$a_i$ for the input $x_i$, the probability that the $N$ players
produce the correct output is thus given by \be\nonumber
P(a_1=x_2,\ldots,a_N=x_1|x_1,\ldots,x_N)=\langle
M_{x_2}^{x_1}\otimes\ldots\otimes M_{x_1}^{x_n}\rangle\,, \ee and
the average  winning probability is \be\label{pq}
\omega=\sum_{\bm{x}} q(\bm{x}) \langle M_{\bm{x}}\rangle\,, \ee
where we have written $M_{\bm{x}}=
M_{x_2}^{x_1}\otimes\ldots\otimes M_{x_1}^{x_n}$ for short.  The
operators $ M_\bm{x}$ satisfy the following properties
\begin{equation}\label{proj}
 M_\bm{x}^2= M_\bm{x}\,,
\end{equation}
and
\begin{equation}\label{orth}
M_\bm{x}M_\bm{y}=0 \quad \text{ if } \bm{x}\neq\bm{y},\bar{\bm{y}}\,.
\end{equation}
The first property follows from the fact that the $M_\bm{x}$ are projection operators. The second property follows from the orthogonality relations $M_{a_i}^{x_i}M_{\bar{a}_i}^{x_i}=0$ and observation ($4$).
Note that protocols involving mixed states or general measurements can all be represented in the above form by expanding the dimensionality of the initial state.

We now show, using (\ref{proj}) and (\ref{orth}), that
$\omega=\sum_\bm{x} q(\bm{x}) M_\bm{x} \leq \omega_c$, where
$\leq$ should be understood as an operator inequality, i.e.,
$A\leq B$ means that $\langle A\rangle\leq \langle B\rangle$ for
all $|\psi\rangle$. First note that  $\sum_\bm{x} q(\bm{x})
M_\bm{x}\leq \sum_\bm{x} q'(\bm{x}) M_\bm{x}$, where $q'(\bm{x})
=q(\bm{x})+(\omega_c-q(\bm{x})-q(\bar{\bm{x}}))/2$ since by
definition $\omega_c-q(\bm{x})-q(\bar{\bm{x}})\geq 0$. It is thus
sufficient to consider weights $q$ such that
$q(\bm{x})+q(\bar{\bm{x}})=\omega_c$ for all $\bm{x}$. We can then
write
\begin{eqnarray}
\omega_c-\sum_\bm{x} q(\bm{x}) M_\bm{x} &=& \left[\sqrt{\omega_c} - \sum_\bm{x} \alpha_\bm{x} M_\bm{x}\right]^2\nonumber\\
&& +\frac{1}{2} \sum_\bm{x} \left[\beta_\bm{x} M_\bm{x} -\beta_{\bar{\bm{x}}} M_{\bar{\bm{x}}}\right]^2\label{sos}
\end{eqnarray}
where $\alpha_\bm{x}=\sqrt{\omega_c}-q(\bar{\bm{x}})/\sqrt{\omega_c}$ and $\beta_\bm{x}=\sqrt{q(\bm{x})q(\bar{\bm{x}})/\omega_c}$. To verify this identity we only need to use (\ref{proj}), (\ref{orth}), and the fact that $q(\bm{x})+q(\bar{\bm{x}})=\omega_c$.  Note now that the right hand-side of (\ref{sos}) is $\geq 0$, since it is a sum of square involving only hermitian operators. This shows that $\sum_\bm{x} q(\bm{x}) M_\bm{x}\leq \omega_c$, as announced.

The inequality $\sum_\bm{x} q(\bm{x})
P(\bm{a}_i=\bm{x}_{i+1}|\bm{x}) \leq \omega_c$ can be interpreted
as a Bell inequality whose local and quantum bound coincide. It is
well known that in order to achieve a Bell violation in quantum
theory one must perform measurements corresponding to
non-commuting operators. The above proof, however, does not distinguish non-commuting operators from ordinary, commuting numbers: it is based on the algebraic identity~(8) which follows only from Eqs.~(\ref{proj}) and (\ref{orth}), regardless of whether the $M_{\bm{x}}$'s commute or not. This explains why the classical and quantum bounds are
identical.

\textbf{GYNI with no-signalling resources}. At first sight, it may
seem that the quantum limitation on the GYNI game arises from the
no-signalling principle:  if the players were able to win with
high probability, their output would somehow depend on their
neighbour's input. This motivates us to look at how players
constrained only by the no-signalling principle perform at GYNI.

Formally, the no-signalling principle states that the marginal distribution $P(a_{i_1},\ldots,a_{i_k}|x_{i_1},\ldots,x_{i_k})$ for any subset $\{i_1,\ldots,i_k\}$ of the $n$ parties should be independent of the measurement settings of the remaining parties~\cite{barrett}, i.e., that
\be\label{no-sign}\nonumber
P(a_{i_1},\ldots,a_{i_k}|x_{1},\ldots,x_{N})=P(a_{i_1},\ldots,a_{i_k}|x_{i_1},\ldots,x_{i_k})
\ee
This guarantees that any subset of the parties is unable to signal to the other by their choice of inputs.

We show in Appendix A that players constrained only by
no-signalling have a bounded winning probability
\linebreak[4]$\omega_{ns}\leq 2\omega_c$. They thus cannot win in
general with unit probability at GYNI.
 Furthermore, for certain input distributions, such as the one where all input strings are chosen with equal weight $q(\bm{x})=1/2^{N}$, we show as expected that $\omega_{ns}=\omega_c$. That is, for uniform and completely uncorrelated inputs, any resource performing better than a classical strategy is necessarily signalling.


Surprisingly, this property is not general. There exist distributions $q(\bm{x})$ for which no-signalling strategies outperform classical and quantum strategies. Consider for instance
the following input distribution
\begin{equation}
\label{promise}
q(\bm{x})=\left\{\begin{array}{ll}
1/2^{N-1} &\text{if } x_1\oplus\cdots\oplus x_{\hat N}=0\\
0 & \text{otherwise}\,,\end{array}
\right.
\end{equation}
where  $\hat N=N$ if $N$ is odd and $\hat N=N-1$ if $N$ is even.
It easily follows from the previous analysis that for  classical
and quantum resources, $\omega_c= 1/2^{N-1}$. We now prove,
however, that no-signalling resources can achieve $\omega_{ns}=
4/3\,\omega_c$. Note that the distribution (\ref{promise}) can be
interpreted as a promise that the sum of the inputs (modulo 2) is
equal to zero. This prior knowledge does not yield any information
to the parties about the value of their neighbour's input, yet it
can be exploited by no-signalling correlations to outperform
classical strategies.

We start by considering the case $N=3$, for which
\begin{eqnarray} \label{3player}
\omega&=&\frac{1}{4}\left[P(000|000)+P(110|011)\right.\nonumber\\
&&\quad+\left.P(011|101)+P(101|110)\right]\,,
\end{eqnarray}
where $P(000|000)=P(a_1=0,a_2=0,a_3=0|x_1=0,x_2=0,x_3=0)$, and so on.
Consider the first three terms in \eqref{3player}. The no-signaling
principle implies that
\begin{eqnarray}\label{3NS bound}
P(000|000)\leq \sum_{a_3}P(00a_3|000)=\sum_{a_3}p(00a_3|001)\,,\nonumber\\
P(110|011)\leq \sum_{a_2}P(1a_20|011)=\sum_{a_2}p(1a_20|001)\,,\\
P(011|101)\leq \sum_{a_1}P(a_111|101)=\sum_{a_1}p(a_111|001)\,.\nonumber
\end{eqnarray}
By normalization of probabilities, the sum of the right-hand sides
of Eqs.~\eqref{3NS bound}  is upper-bounded by one, and thus
$P(000|000)+P(110|011)+P(011|101)\leq 1$. Similar conditions are
obtained for any of the four possible combination of three terms
in Eq.~\eqref{3player}. Summing over these possibilities, we find
$3[P(000|000)+P(110|011)+P(011|101)$$+P(101|110)]\leq 4$, or in
other words $\omega_{ns}\leq 4/3\times 1/4=4/3\,\omega_c$.
Furthermore the inequality is saturated only if the four
probabilities appearing in \eqref{3player} are all equal to $1/3$.
It turns out that the remaining entries of the probability table
$P(\bm{a}|\bm{x})=P(a_1a_2a_3|x_1x_2x_3)$ can be completed in a
way that is compatible with the no-signalling principle, i.e, the
bound $\omega_{ns}\leq 4/3\,\omega_c$ is achievable. Up to
relabeling of inputs and outputs, there exist two inequivalent
classes of extremal no-signalling correlations achieving this
winning probability (see Appendix B). One of them takes the form
$P(\bm{a}|\bm{x})=2/3\,g(\bm{a},\bm{x})+ 1/3\,g'(\bm{a},\bm{x})$
where $g$ and $g'$ are the following boolean functions
\begin{equation}
\begin{split}
g(\bm{a},\bm{x})= &a_1 a_2 a_3 (1 \oplus x_1) (1 \oplus x_2) (1 \oplus x_3)\\
g'(\bm{a},\bm{x})=& (1 \oplus a_1)(1 \oplus a_2)(1 \oplus a_3) \\
& \oplus x_1a_2a_3 \oplus a_1x_2a_3 \oplus a_1a_2x_3 \oplus x_1x_2x_3\,.
\end{split}
\end{equation}
From this definition, it is easy to verify that $P(a_1a_2a_3|x_1x_2x_3)$ satisfies the no-signalling conditions and achieves winning probability $\omega_{ns}=1/3=4/3\,\omega_c$.

The existence of no-signaling correlations achieving
$\omega_{ns}=4/3\, \omega_c$ in the case $N=3$ is enough to show
that $\omega_{ns} \geq 4/3\, \omega_c$ for any $N\geq 3$. This can
be seen as follows. Consider the situation in which the first
three parties use the optimal strategy for $N=3$ while the
remaining parties simply output their input. In this case, all the
terms in $\omega$ vanish, except the four terms $P(000,0\ldots
0|000,0\ldots 0)$, $P(110,0\ldots 0|011,0\ldots 0)$,
$P(011,1\ldots 1|101,1\ldots 1)$, and $P(101,1\ldots 1|110,1\ldots
1)$, which are all equal to $1/3$.

Beyond these analytical results, we obtained the maximal
no-signaling values of $\omega_{ns}$ up to $N=7$ players using
linear programming. The ratios $\omega_{ns}/\omega_{c}$ of
no-signalling to classical winning probabilities are $4/3$ for
$N=3,4$, $16/11$ for $N=5,6$, and $64/42$ for $N=7$, showing that
for more parties there exist no-signaling correlations that can
outperform the optimal no-signaling strategy for $N=3$. (Note that
it can be shown that the winning probability for an odd number $N$
of parties is always equal to the winning probability for $N+1$
players, see Appendix C).

\textbf{GYNI Bell inequalities.} The GYNI Bell inequalities
$\sum_{\bm{x}} q(\bm{x}) P(\bm{a}_i=\bm{x}_{i+1}|\bm{x}_i)\leq
\omega_c$ are not violated by quantum theory, but can be violated
by more general no-signalling theories. In \cite{Prob26Werner},
Gill raised the question of whether there exist Bell inequalities which
(i) feature this `no quantum advantage' property and (ii) define
facets of the polytope of local correlations. Here we give a
positive answer to this question. We have checked that the GYNI
inequalities  defined by the distribution \eqref{promise} are
facet-defining for $N\leq 7$ players. More generally, we
verified that the inequalities defined  by the distribution
$q(\bm{x})$ having uniform support on $\bigoplus_{i=1}^{\hat
N}x_i=0$ are facet-defining for all $N\leq 7$. We conjecture that
they are facet-defining for any number of parties. Note also that
the polytope of local correlations for the case $N=3$ (with binary
inputs and outputs) was completely characterized in \cite{sliwa};
the inequality corresponding to (\ref{3player}) belongs to the
class 10 of \cite{sliwa}. Geometrically, our result shows that the
polytope of local correlations and the set of quantum correlations
have in common faces of maximal dimension (we recall that a facet
corresponds to a $(d-1)$-dimensional face of a $d$-dimensional
polytope).

This also implies that GYNI is an information-theoretic game that
identifies a portion of the boundary of quantum correlations which
is of non-zero measure. To the best of our knowledge, all
previously introduced information-theoretic or physical principles
recovering part of the quantum boundary -- including nonlocal
computation \cite{NLC}, nonlocality swapping \cite{emergence},
information causality \cite{IC,allcock}, and macroscopic locality
\cite{ML} -- only single out a portion of zero-measure
\cite{note}.

\textbf{Discussion and open questions.} Our work raises plenty of
new questions. First, it would be interesting to understand the
structure of those input distributions $\jd$ leading to a gap
between no-signaling and classical/quantum correlations (See
Appendix A, for a class of distributions for which there is no gap). For
instance, in the case of four parties, the distribution $\jd$
having uniform support on $x_1\oplus x_2 \oplus x_3 \oplus
x_1x_2x_3=0$ leads to $\omega_{ns}=4/3\, \omega_c$. However, the
corresponding Bell inequality is not a facet. Another question is
thus to single out, among all relevant input distributions, those
corresponding to facet Bell inequalities. For three parties,
it follows from \cite{sliwa} that the distribution
\eqref{promise} is the unique possibility.

A further interesting problem is whether there exist facet Bell
inequalities  with no quantum advantage in the bipartite case.
Note that our GYNI inequalities are non-trivial only for $N\geq
3$; for the case $N=2$, the classical and no-signalling bounds are
always equal. In ref.~\cite{NLC}, examples of bipartite Bell
inequalities with no quantum advantage have been presented in the
context of nonlocal computation. However, as mentioned earlier,
none of the Bell inequalities associated to nonlocal computation
has been proven to be facet-defining. We studied this question
here and could prove that none of the simplest inequalities from
\cite{NLC} (corresponding to the family of inequalities specified by the parameters $n=2,3$ in \cite{NLC}) are facet inequalities. The proof uses a mapping from
these inequalities to the space of correlation inequalities for
$n$ parties, two settings and two outcomes, which was fully
characterized in Ref.~\cite{werner&wolf}; see Appendix D for a
detailed proof. We conjecture that none of the Bell inequalities introduced in
\cite{NLC} are facet-defining.

Coming back to our original motivation, it would be interesting to
get a deeper understanding of the structure and
information-theoretic properties of the no-signaling correlations
giving an advantage over classical/quantum correlations, for
instance those associated to inequality~(\ref{3player}). In
particular, it would be interesting to understand if they can be
exploited for other information tasks. Finally, our results
suggest that the quantum limitation on the GYNI game might
originate from a generalization of the no-signalling principle in
a multipartite setting. Can this intuition be made concrete? Are
there more general information tasks with no quantum advantage?

 \textit{Acknowledgments.} We thank C. Branciard and D. Perez-Garcia for discussions,
and the QAP Partner Exchange Programme. This work is financially
supported by the Funda\c{c}\~{a}o para a Ci\^{e}ncia e a
Tecnologia (Portugal) through the grant SFRH/BD/21915/2005, the
European ERC-AG Qore and PERCENT projects, the Spanish MEC FIS2007-60182
and Consolider-Ingenio QOIT projects, Generalitat de Catalunya and
Caixa Manresa, the UK EPSRC, the Interuniversity Attraction Poles (Belgian Science Policy) project IAP6-10 Photonics@be, the EU project QAP contract 015848, and the Brussels-Capital region through a BB2B grant.

\bibliographystyle{plain}

\begin{thebibliography}{99}
\bibitem{note} Note that Ref. \cite{ML} can actually recover the entire
quantum boundary in the correlator space.
\end{thebibliography}

\newpage

\section{Appendix A}
Here we derive the upper bound $\omega_{ns}\leq 2\omega_{c}$ for the winning probability $\omega_{ns}$ of no-signalling strategies. We then show that $\omega_{ns}=\omega_c$ for all input distributions $\jd(\bf{x})$ such that $q(\bm{x})\leq q(\bm{y})=q(\bm{\bar y})$ for some input string $\bm{y}$. Such distributions include in particular the uniform distribution where all input strings are chosen with equal weight $q(\bm{x})=1/2^N$.

To start we derive the upper bound $\omega_{ns}\leq 2\omega_c$, valid for any
distribution $\jd(\bf{x})$. From the definition~(\ref{cstrat}), we have that
$\jd(\bf{x})\leq\omega_c$ for every input string $\bf{x}$. This trivially leads to the upper-bound
\begin{equation}\label{omega bound}
\omega_{ns}\leq \omega_c\sum_{\bm{x}} P(\bm{a}_i=\bm{x}_{i+1}|\bm{x})\,.
\end{equation}
Notice that this bound is only meaningful when the right-hand side
is smaller than 1, since obviously $\omega_{ns}\leq1$.
We now show that for all no-signalling distributions $\sum_{\bm{x}} P(\bm{a}_i=\bm{x}_{i+1}|\bm{x})\leq 2$, from which the bound $\omega_{ns}\leq 2\omega_c$ immediately follows.

First note that from the no-signaling condition,
\be\label{rec}
P(a_{1},\ldots,a_{k-1}|x_{1},\ldots,x_{k-1})\leq P(a_{1},\ldots,a_{k}|x_{1},\ldots,x_{k})\,.
\ee
We now write
\begin{eqnarray}
&&\sum_{\bm{x}} P(\bm{a}_i=\bm{x}_{i+1}|\bm{x})\nonumber\\
&&=\sum_{x_1,\ldots,x_N} P(a_1=x_2,\ldots,a_N=x_1|x_1,\ldots,x_N)\nonumber\\
&&\leq \sum_{x_1,\ldots,x_N}P(a_1=x_2,\ldots,a_{N-1}=x_N|x_1,\ldots,x_{N-1})\nonumber\\
&&= \sum_{x_1,\ldots,x_{N-1}} P(a_1=x_2,\ldots,a_{N-2}=x_{N-1}|x_1,\ldots,x_{N-2})\,\nonumber
\end{eqnarray}
where the inequality follows from \eqref{rec} and in the last equality we used the no-signaling condition after summing over $x_N$.
Iteratively performing this last step, we finally obtain
\be
\sum_{\bm{x}} P(\bm{a}_i=\bm{x}_{i+1}|\bm{x})\leq \sum_{x_1,x_2} P(a_1=x_2|x_1)\leq2\,.
\ee

We now analyze the no-signalling winning probability for distributions satisfying $q(\bm{x})\leq q(\bm{y})=q(\bm{\bar y})$ for some input string $\bm{y}$. Note that for such weights $\omega_c=q(\bm{y})+q(\bm{\bar y})=2q(\bm{y})$, as easily follows from (\ref{cstrat}). We thus have
\be
\omega_{ns}=\sum_{\bm{x}}q(\bm{x}) P(\bm{a}_i=\bm{x}_{i+1}|\bm{x})\nonumber
\leq \frac{\omega_c}{2}\sum_{\bm{x}}P(\bm{a}_i=\bm{x}_{i+1}|\bm{x})\,.
\ee
But, as we have shown above, $\sum_{\bm{x}} P(\bm{a}_i=\bm{x}_{i+1}|\bm{x})\leq 2$ for all no-signalling distributions , and thus $\omega_{ns}\leq\omega_c$. Since any classical strategy is also a no-signalling strategy, it actually holds that $\omega_{ns}=\omega_c$.

\section{Appendix B}

Here we describe two inequivalent no-signaling correlations which
attain $\omega_{ns}=4/3\ \omega_c$ for the tripartite inequality
\eqref{3player}. These correlations are extremal non-local boxes
in the sense of being vertices of the no-signaling polytope for
three parties and binary inputs and outputs~\cite{barrett}.


Writing $(a,b,c)$ for $(a_1,a_2,a_3)$ and $(x,y,z)$ for $(x_1,x_2,x_3)$, we can write
the first box as
\begin{equation}
P_1(a,b,c|x,y,z) = \frac13 f(a,b,c,x,y,z)
\end{equation}
where $f(a,b,c,x,y,z)$ is the boolean function
\begin{equation}
\begin{split}
f(a,b,c,x,y,z) = &(1 \oplus b \oplus x \oplus y \oplus xy)(1 \oplus c \oplus z)\\
&\oplus a(1 \oplus y \oplus cy \oplus b(c \oplus z)).
\end{split}
\end{equation}
Similarly, we can write the second box as
\begin{equation}
P_2(a,b,c|x,y,z) = \frac23 g(a,b,c,x,y,z) + \frac13 g'(a,b,c,x,y,z)
\end{equation}
with $g$ and $g'$ the two boolean functions
\begin{equation}
\begin{split}
g(a,b,c,x,y,z)=& a b c (1 \oplus x) (1 \oplus y) (1 \oplus z)\\
g'(a,b,c,x,y,z)=& (1 \oplus a)(1 \oplus b)(1 \oplus c) \\
& \oplus xbc \oplus ayc \oplus abz \oplus xyz\,.
\end{split}
\end{equation}
Among the boxes that are equivalent to $P_1$ under relabeling of
parties, inputs, and outputs, a total of 24 of them  violate
maximally the Bell inequality \eqref{3player}, and similarly for 8
of those that are equivalent to $P_2$. Even though other
tripartite no-signaling boxes (inequivalent to $P_1$ or $P_2$
under relabeling of parties, inputs, or outputs) violate the Bell
inequality \eqref{3player}, those 32 boxes obtained from $P_1$ and
$P_2$ are the unique ones that violate it maximally.

\section{Appendix C}
Here we show that for the input distribution \eqref{promise}, the
no-signaling bound for an even number of parties $N+1$ is always
equal to the no-signaling bound for $N$ parties. Start by
considering $N+1$-GYNI game, where the first $N$ players use the
optimal strategy for the $N$-player case and player $N+1$ outputs
its input. They then achieve a no-signaling violation equal to the
$N$ case, which imposes the lower bound
$\omega_{ns}(N+1)\geq\omega_{ns}(N)$. But this is actually the
best average success these $N+1$ players can obtain. To see that,
consider the game for $N+1$ parties. Allowing players $N$ and
$N+1$ to communicate can only increase the achievable value of
$\omega_{ns}(N+1)$. Indeed, in this situation the best strategy
that player $N$ can adopt is to output $x_{N+1}$, which was
communicated to him by player $N+1$, while player $N+1$ needs to
guess $x_1$ given $x_N$ and $x_{N+1}$. Clearly, the knowledge of
$x_{N+1}$ is of no use for him since this bit is completely
uncorrelated with the rest of the input string. Consequently, the
situation is analogous to having players $1,\ldots,N-1,N+1$ (i.e.
all players except player $N$) play the game for $N$ parties.
Therefore $\omega_{ns}(N+1)\leq \omega_{ns}(N)$ and we have
finally that $\omega_{ns}(N+1)=\omega_{ns}(N)$ for odd $N$.

\section{Appendix D}
In what follows, we derive a criterion that is necessarily satisfied by any facet-defining Bell inequality associated to the task of nonlocal computation (NLC) \cite{NLC}, and show that none of the NLC Bell inequalities for boolean functions of two and three input bits are facet-defining.

Nonlocal computation is a distributed task of two parties, where
the goal is to compute a given boolean function $f(\textbf{z})$ of
an $n$-bit string $\textbf{z}$. The input bit string is decomposed
into two strings $\textbf{x}$ and $\textbf{y}$, such that
$\textbf{x}\oplus \textbf{y}=\textbf{z}$. The bit string
$\textbf{x}$ is sent to party A while the bit string $\textbf{y}$
is sent to party B. Upon receiving their input bit strings, A and
B each output a single bit, $a$ and $b$ respectively, such that
the following relation holds: $a\oplus b=f(\textbf{z})$.
Importantly, each party has locally no information about the input
bit string $\textbf{z}$, that is $P(x_i=z_i)=1/2$ for all
$i=1,...,n$. For each $n$, $f(\textbf{z})$, and distribution of
inputs $\tilde p(\textbf{z})$, we obtain a Bell expression whose
value is associated to the probability of success at the task.
These NLC inequalities have the form
\begin{equation} \label{LPSW}
I(n,f,\tilde p)=\sum_{\textbf{z}}(-1)^{f(\textbf{z})}\tilde p(\textbf{z})\sum_{\textbf{x}\oplus
\textbf{y}=\textbf{z}}\ev{A_xB_y}\leq k(n,f,\tilde p)
\end{equation}
where $A_x$ and $B_y$ are observables which take values
$\{-1,1\}$. Notice that each party measures $2^n$ observables.

In Ref.~\cite{NLC} it is proven that the best classical strategy is given by
$A_x=(-1)^{a_x}$ and $B_y=(-1)^{b_y}$ with
\begin{equation}\label{opt strat}
a_x=\textbf{u} \cdot \textbf{x}, \qquad b_y=\textbf{u}\cdot\ \textbf{y} \oplus \delta\,,
\end{equation}
where $\delta$ denotes a single bit and $\textbf{u}$ an $n$-bit
string shared by the parties. This classical strategy, which is a
linear approximation of the function $f$, achieves a winning
probability as high as any quantum resource. Thus the local and
quantum bounds of inequalities \eqref{LPSW} coincide. There exist
however no-signaling correlations which can perform with winning
probability one at this game.

Checking whether the NLC inequalities \eqref{LPSW} are facet-defining is in
general a hard problem since one should consider any
input size $n$, boolean function $f$, and distribution $\tilde
p(\textbf{z})$. Below we give a first simplification to this problem by deriving a necessary criterion satisfied by facet NLC inequalities. Our method is based on a mapping from the $(2,2^n,2)$ correlation space --  i.e. (2 parties, $2^n$
settings, 2 outcomes) -- in which the NLC inequalities are defined, into the
$(n,2,2)$ full-correlation space for which the complete set of
tight Bell inequalities has been provided in
Ref.~\cite{werner&wolf}.

To any inequality of the form (\ref{LPSW}) defined by the triple $(n,f,\tilde p)$, we associate the following Bell inequality in the $(n,2,2)$ full-correlation space:
\begin{equation} \label{LPSWncorr}
I_{n22}(n,f,\tilde p)=\sum_{\textbf{z}}c(\textbf{z})\ev{C_{z_1}\ldots C_{z_n}}\leq
2^{-n}k(n,f,\tilde p)
\end{equation}
where $c(\textbf{z})=(-1)^{f(\textbf{z})}\tilde p(\textbf{z})$, and where we view $z_i\in\{0,1\}$ as denoting one of two possible observables $C_{z_i}$ of party~$i$ taking values $\{-1,1\}$ (with $i=1,\ldots,n$).

\begin{lemma}If the NLC inequality $I(n,f,\tilde p)$ for $n$ bits is facet-defining, then the corresponding inequality $I_{n22}(n,f,\tilde p)$ is facet-defining in the $(n,2,2)$ full-correlation space.\end{lemma}
\emph{Proof.}
The deterministic extremal points of the $(n,2,2)$ full-correlation polytope are of the form~\cite{werner&wolf}
\be\label{det}
\langle C_{z_1}\ldots C_{z_n}\rangle=(-1)^{u_1z_1}\ldots(-1)^{u_n z_n}(-1)^{\delta}=(-1)^{\textbf{u}\cdot \textbf{z} \oplus\delta}
\ee
where $u_i\in\{0,1\}$ specifies the local strategy of each party and $\delta\in\{0,1\}$ represents an additional global sign flip, which we can think of as being carried out by the last party. These deterministic points are thus specified by the single bit $\delta$ and the $n$-bit string $\textbf{u}$, and are therefore in one-to-one correspondence with the extremal points (\ref{opt strat}) saturating the inequalities (\ref{LPSW}). For any such strategy specified by $\delta$ and $\bm{u}$, we have that
\be\begin{split}
\sum_{\bm{x}\oplus\bm{y}=\bm{z}}\langle A_xB_y\rangle&=\sum_{\bm{x}\oplus\bm{y}=\bm{z}}(-1)^{\textbf{u}\cdot (\bm{x}+\bm{y}) \oplus\delta}\\
&=2^n (-1)^{\textbf{u}\cdot \textbf{z} \oplus\delta}=2^n \langle C_{z_1}\ldots C_{z_n}\rangle\,.
\end{split}\ee
It immediately follows from the above identity that the inequalities (\ref{LPSWncorr}) are valid for the $(n,2,2)$ full-correlation polytope.

Let us now suppose that the Bell inequality $I_{n22}(n,f,\tilde
p)\leq 2^{-n}k(n,f,\tilde p)$ is not facet-defining. Then we can
write $I_{n22}(n,f,\tilde p)=I^1_{n22}(n,f,\tilde
p)+I^2_{n22}(n,f,\tilde p)$ and $k(n,f,\tilde p)=k^1(n,f,\tilde
p)+k^2(n,f,\tilde p)$ for some $I^1_{n22}(n,f,\tilde p)$,
$I^2_{n22}(n,f,\tilde p)$, $k^1(n,f,\tilde p)$, and
$k^2(n,f,\tilde p)$ such that \be I^1_{n22}(n,f,\tilde p)\leq
2^{-n}k^1(n,f,\tilde p) \ee and \be I^2_{n22}(n,f,\tilde p)\leq
2^{-n} k^2(n,f,\tilde p) \ee are valid inequalities for the
$(n,2,2)$ full-correlation polytope, i.e., they are satisfied by
all deterministic points of the form (\ref{det}). But then it
follows from the above correspondence between deterministic point
of the $(n,2,2)$ polytope and the $(2,2^n,2)$ polytope that
$I(n,f,\tilde p)=I^1(n,f,\tilde p)+I^2(n,f,\tilde p)$, where \be
I^1(n,f,\tilde p)\leq k^1(n,f,\tilde p) \ee and \be I^2(n,f,\tilde
p)\leq k^2(n,f,\tilde p) \ee are valid NLC inequalities. This
implies that $I(n,f,\tilde p)\leq k(n,f,\tilde p)$ is not
facet-defining for the $(2,2^n,2)$ polytope, from which the
statement of the Lemma follows. \hfill $\square$

The above Lemma implies that it is sufficient to restrict our analysis on NLC inequalities associated with facet inequalities in the $(n,2,2)$-full correlation space.
In Ref.~\cite{werner&wolf} a construction for the coefficients $c(\textbf{z})$ of all facet $(n,2,2)$ correlation Bell inequalities has been given. For small number of inputs, i.e. $n=2,3$, we have explicitly verified that none of the corresponding NLC inequalities are facet-defining; all these inequalities can actually be expressed as sums of CHSH inequalities. For larger $n$ however, a similar analysis becomes difficult due to the large number of facet $(n,2,2)$ Bell inequalities and the high dimensionality of the $(2,2^n,2)$ correlation space.
\end{document}